**Highly stable, reactive and ultrapure nanoporous metallic films**

*Hyunah Kwon, Hannah-Noa Barad, Alex Ricardo Silva Olaya, Mariana Alarcon-Correa, Kersten Hahn, Gunther Richter, Gunther Wittstock, Peer Fischer \**

H. Kwon, H. Barad, M. Alarcon-Correa, G. Richter
Max Planck Institute for Intelligent Systems, Heisenbergstrasse 3, 70569 Stuttgart, Germany

A. R. Silva Olaya, G. Wittstock
School of Mathematics and Science, Department of Chemistry, Carl von Ossietzky University of Oldenburg, 26111 Oldenburg, Germany

K. Hahn
Max Planck Institute for Solid State Research, Heisenbergstrasse 1, 70569 Stuttgart, Germany

P. Fischer
Institute for Molecular Systems Engineering, Heidelberg University, INF 253, 69120 Heidelberg, Germany
E-mail: peer.fischer@mr.mpg.de

**Abstract**

Nanoporous metals possess unique properties attributed to their high surface area and interconnected nanoscale ligaments. They are mostly fabricated by wet synthetic methods involving solution-based dealloying processes whose purity is compromised by residual amounts of the less noble metal. Here, we demonstrate a novel dry synthesis method to produce nanoporous metals, which is based on the plasma treatment of metal nanoparticles formed by physical vapor deposition. Our approach is general and can be applied to many metals including non-noble ones. The resultant nanoporous metallic films are impurity-free and possess highly curved ligaments and nanopores. The metal films are remarkably robust with many catalytically active sites, which is highly promising for electrocatalytic applications.

## 1. Introduction

Nanoporous metals (NPMs) exhibit unique properties distinct from bulk metals.[1] The nanoscale pore-architecture directly determines the optical,[2] electrical,[3] chemical,[4] and mechanical[5] properties. The high surface-area-to-volume ratios and large number density of high surface energy atoms at curved surfaces give rise to significant catalytic activities that are absent in the bulk metal.[6,7] NPMs are also lightweight and can be optically transparent.[8,9] Interconnected films of NPMs have thus attracted strong interest for applications as electrode materials for synthesis,[10] energy conversion[11] and sensing,[12] or as plasmonic materials.[13] Since NPMs do not occur naturally, they need to be synthesized.

Many different schemes to synthesize NPMs have been developed.[14–17] A few studies report the deposition of metals onto pre-formed inorganic or polymeric nanoporous templates, typically by electrochemical deposition or sputtering, followed by an etching step that removes the sacrificial porous support structure.[18–21] However, the requirement of a separate template and multiple fabrication steps is a disadvantage. A wet-chemical sol-gel scheme that relies on the reduction of metal salts to form metal nanoparticles followed by their aggregation, has also been reported.[8,15] While it affords some freedom to vary the composition, the resulting network shows few highly curved nanoporous features and thus few grain boundaries and defect sites at the surface.[22] Highly curved nanoporous features are desirable, as they support low-coordinated atoms that are favorable sites for catalytic activity.[23] By far, the most widely used method to synthesize NPMs is dealloying as it can yield reactive metallic films with higher reactivity than the flat homologue. Dealloying involves the selective chemical etching of one or more less noble metals out of an alloy.[14,24,25] For instance, nanoporous Au structures are generally formed by etching silver out of Ag-Au alloys.[26] While the method is relatively simple and effective, the less noble metal, here Ag, cannot be completely removed for thermodynamic reasons. The remaining content of the less noble element is difficult to precisely control and may greatly affect the catalytic activity of the resulting NPM.[27,28] For brevity we call the remaining content of the less noble metal in the NPM an "impurity". Another challenge is that the dealloying process to form NPMs with non-noble metals requires carefully chosen combinations of materials and chemistries, so that the metals in the master alloy can be selectively etched. Generalizing the fabrication conditions using a wet synthesis is thus challenging.[29,30] It is therefore desirable to develop a scheme that permits catalytically active NPMs to be formed from several metals, including non-noble metals, and free from impurities.

Here, we demonstrate a scheme to obtain impurity-free NPM films (NPMFs) with highly curved pore structures of adjustable sizes. The NPMFs are formed by the coalescence of metal nanoparticles in a low-temperature plasma. No solution-processing or harsh chemicals are required, and the process can be generally applied. We form nanoporous Au, Ag, Pt, Pd, Ni, and Fe films. Our method yields pure nanoporous Au without any secondary metal impurities. We demonstrate methanol oxidation and the oxygen evolution reaction (OER) with our nanoporous Au film (NPGF) and Ni film (NPNF), respectively. Our NPMFs are catalytically highly active. Remarkably, electron microscopy and the catalytic reactivity both indicate the presence of high surface energy sites in our NPGF that are observed to be structurally more robust than those found in dealloyed nanoporous Au.

## 2. Results and Discussion

### 2.1. Dry synthesis of mesoporous metal-films

Our dry synthesis of NPMFs is based on the plasma treatment of a dense layer of nanoparticles deposited on a sacrificial thin polymer film. The process is schematically depicted in Fig. 1a. First, a 1 μm thick PMMA film is formed on a flat substrate, such as silicon, by spin-coating (see methods). A dense layer of metal nanoparticles with a diameter of about 10 nm is then obtained by e-beam evaporation of the metal onto the dried PMMA film at an oblique angle, here 80° (see Supplementary Fig. 1). The deposited metal atoms (*e.g.*, Au, Ag, Pt, Pd, Ni, Fe) form random clusters on the substrate, which give rise to shadowing under oblique angle growth. This growth results in a rough nanostructured thin-film formed by a dense layer of metallic nanoparticles on top of the PMMA film. The substrate is then treated in a simple laboratory plasma system. During the plasma treatment, the PMMA starts to be etched by plasma ions that penetrate through the gaps between the metal nanoparticles (Fig. 1b1). As a result, the PMMA layer becomes rough and shrinks, as schematically shown in Fig. 1b2. The metal nanoparticles remain compact on the shrinking PMMA islands since the interfacial energy is small. At the same time, the metal nanoparticles are also bombarded with high-energy ions, which renders them mobile and causes their coalescence,[31] facilitating cluster migration at low temperature (Fig. 1b3).[32] We presume that the thermal energy is not high enough to liquefy the nanoparticles and cause their fusion, but rather that the coalescence leads to polycrystalline structures with many grain boundaries (see transmission electron microscope analysis below).[33,34] When the remaining PMMA layer has been completely etched away, the porous metal structures will contact the substrate, where the metal is less mobile, since the interfacial energy is lower compared to that between nanoparticles and PMMA. Coalescence then ceases and the NPMF remains robust despite further plasma treatment (Fig. 1b4, Supplementary Fig. 2). Note that shadow growth is necessary to obtain a continuous nanoporous film via plasma treatment as is indicated in Supplementary Fig. 3.

------FIGURE 1-------

Heating the nanoparticle-decorated PMMA film without plasma treatment did not yield nanoporous metallic structures (Supplementary Fig. 4). Hence, the process is not simply described by thermal effects. Rather, the etching of the PMMA plays an important role, in combination with surface tension effects and the aforementioned differences in interfacial energies between the metal and the two substrates. Fig. 1c shows scanning electron microscope (SEM) images of the resulting mesh-like NPMFs fabricated with Au, Pt, Pd, and Ag, and the non-noble metals Ni and Fe. The films have pore sizes between 20 nm and hundreds of nanometers, and the ligament sizes of the metal ligaments range from 10 nm to 80 nm. An exemplary ligament in the network is marked by red lines in Fig. 1c. The ligament or pore sizes can be tuned by controlling the conditions during the plasma treatment. For example, the nanoporous Au film (NPGF) can have ligament sizes from 10 to 30 nm depending on the plasma time (Supplementary Fig. 6). The nanoporous Au and Ni morphology can be controlled through the gas composition and the plasma power (Supplementary Fig. 7). The kinetic energy of the ions in the plasma will vary according to the plasma conditions, resulting in different metal atom mobilities and different etching rates of the sacrificial PMMA layer.

We assume that the NPMF morphology is mainly a function of the melting points of the metals, surface energies of the metals, and the interfacial energies between the PMMA and metals. The interfacial energies are difficult to measure or predict. Nevertheless, a trend emerges by comparing different morphologies of various metals and their respective melting points and surface energies, as shown in Supplementary Table 1 and Fig. 8. For example, Ag has the lowest melting point and surface energy of the metals in the table and it gives rise to a NPMF that shows a small variation in the pore and ligament sizes. This may be explained by the low surface energy of Ag which means that Ag NPs will not aggregate as much as the other metals. In contrast, Pt has the highest melting point and surface energy in the table and it gives rise to non-uniform morphology with larger variations in pore and ligament sizes. This can be because Pt NPs are less mobile and tend to aggregate quickly, compared to Ni or Fe NPMFs.

We checked that the nanoporous networks are fully connected over large areas by image processing (pixel connectivity), as is seen in the connectivity color-map (Supplementary Fig. 9a). Supplementary Fig. 9b confirms that >99% of ligaments are very well-connected. Also, the sheet resistance of the NPGF was determined to lie in the range of 150 ~ 300 Ω/sq depending on the film morphology (4 probe measurement with a separation of ~ 1.3 mm between probes), which confirms the high connectivity of the NPMF.

## 2.2. Purity of NPGF

A combustion analysis was carried out to check for carbon residues in the NPGF. In this method, $CO_2$ resulting from the combustion of carbon present in the sample can be detected by infrared absorption.[35] Only 5 ppm (wt.) carbon residues from the plasma treatment could be detected. Hence, our NPMFs do not contain any significant carbon residues from the plasma treatment. This is also confirmed in the subsequent X-ray photoelectron spectroscopy (XPS) and electrochemical analysis. As a note, the solubility of C in solid Au is known to reach only 50 ppm at temperatures above 1000 °C.

----- FIGURE 2 -------

We also carried out XPS measurements to evaluate the surface of the NPGF and its purity. The Au 4f signals from an NPGF on a Si wafer are plotted in Fig. 2a. The signals indicate a typical metallic Au $4f_{7/2}$ binding energy of 84.4 eV with the Au $4f_{5/2}$ component separated by a spin-orbit splitting of 3.7 eV.[36] The Au $4f_{7/2}$ component at 84.1 eV is known as the surface component of Au,[37] and is consistent with the large surface area of the NPGF. No peaks are seen that correspond to $Au_2O_3$.[38] The O 1s and C 1s spectra from the NPGF and a spin-coated PMMA layer are compared in Fig. 2b and 2c, respectively. The O 1s spectrum only shows one symmetric peak with a maximum at 533.5 eV which corresponds to SiO$_2$ of the Si substrate.[36] It is known that the O 1s peak for the metal oxide (excluding SiO$_2$) appears below 531 eV, and it is clearly absent in our NPGF. The characteristic O 1s spectrum from a film containing only spin-coated PMMA[39] disappears in the NPGF sample, indicating that PMMA is not present in the NPGF. The C 1s spectrum of PMMA in Fig. 2c shows multiple peaks at around 288.9, 286.9, 285.5, and 283.5 eV that originate from O–C=O, C–O, C–C=O, and –CH$_3$ bonds, respectively.[39] However, these peaks do not appear in the NPGF sample (besides a very low intensity peak, which may relate to C–O state), which corroborates that the PMMA is completely removed by the plasma. The overall C content is drastically reduced in the NPGF sample and originates from unavoidable contamination during sample transfer. Our combustion analysis and XPS results indicate that the PMMA is completely etched away and that no significant C impurities from PMMA remain after the plasma etching step. Compared to nanoporous Au prepared by dealloying or solution phase synthesis, the dry synthesis reported here is a much more convenient and general route to obtain NPMFs that are also free from impurities.

### 2.3. TEM characterization of NPGF

Transmission electron microscopy (TEM) measurements were performed on the NPGF. Fig. 3a shows that the diameter and length of the ligaments in the NPGF range from 10 to 20 nm, and from 10 to 30 nm, respectively, and are relatively small compared to other reported nanoporous Au structures. Numerous pores are observed and the surface shows a high curvature including many concave and protruded structures, as well as many grain boundaries that are found on each curved region. One of the protruded sections in the image was also investigated by energy dispersive X-ray (EDX) mapping, as shown in Fig. 3b. The image shows that the Au is uniformly distributed over the NPGF structure. Moreover, C and O maps are uniform and indistinguishable from the background signals, in agreement with our previous observation that the NPGFs do not contain any significant carbon or oxygen contamination from the sacrificial PMMA.

----- FIGURE 3 ------

High-resolution TEM (HR TEM) images revealing the atomic structure and crystallographic surface facets are shown in Fig. 3c. Fast Fourier transformation (FFT) images were also obtained in the regions of (i) ~ (iv) as depicted in Fig. 3c, clearly showing the polycrystallinity of the NPGF. The grains have a size of a few nanometers that is comparable to the initial size of metal nanoparticles. The measured lattice spacing in each region in the HRTEM image are 0.91, 2.35, 1.44, and 2.35 Å, which correspond to the crystallographic orientations (420), (111), (220), and (111), respectively. This implies that our NPGF has various surface symmetries with multiple directions on the curved ligaments, even if the actual surface facets cannot be determined due to the projection effect in TEM. The polycrystalline nature of our NPGF is also confirmed by X-ray diffraction (XRD) measurements, as shown in Supplementary Fig. 10. The NPGF clearly shows Au polycrystalline structures, with higher peak intensities than the as-deposited Au nanoparticles before plasma treatment indicating higher crystallinity of the NPGF. For the Au nanoparticles before plasma treatment an $Au_2O_3$ (311) peak also appears, which is not present after plasma treatment. This change is attributed to restructuring by bombardment with high-energy ions that removes the native oxide.

One can also clearly identify many grain boundaries between the nanograins. They are formed when the nanoparticles approach each other and aggregate together during plasma treatment,[40,41] resulting in rather straight grain boundaries with thermal grooves having higher energy of migration. It is known that these straight grain boundaries have a much smaller migration velocity compared to the curved ones.[42] Therefore, higher energy is required for the diffusion of metal atoms across the grain boundary compared to diffusion along a crystal facet.[34] The observed high density of straight grain boundaries is a special feature of this type of nanoporous metal and explains the structural robustness of our NPGF. Most curved areas have various surface structures as investigated in other parts of the mesh structure (Supplementary Fig. 11). Multiple surface directions and many grain boundaries that have high surface energies and potentially high-indexed surfaces are ideal sites for catalytic reactions, as we shall confirm next.

## 2.4. Electrochemical properties of NPGF

To explore the catalytic properties of the NPMFs we chose to investigate the catalytic properties of the NPGF using electrochemical measurements. Cyclic voltammograms (CVs) were recorded with nanoporous Au powder recovered from nanoporous gold films (NPGF powder) in acidic and in alkaline solutions. Figure 4a shows the CVs of NPGF powder in 0.1 M $H_2SO_4$ at a scan rate of

10 mV s$^{-1}$. For these measurements, cavity microelectrodes (CMEs) were filled with NPGF powder. In the positively going scan, surface oxidation takes place from 1.2 V showing a very broad anodic peak with a maximum at 1.47 V. The broad peak is attributed to the rough and disordered Au surfaces.[43] A symmetric peak centered near 1.15 V is observed in the negatively going scan, which is associated with the reduction of the Au oxides formed during the preceding positively going scan. Notably, our NPGF powder does not undergo significant surface reconstruction even after 200 cycles, as shown in Fig. 4a. The peaks for the single crystal domains in the potential range of surface oxidation are still broad and poorly defined. This is in contrast to the behavior of dealloyed nanoporous Au, which gives rise to surface restructuring and CVs that change significantly during the initial potential cycles in 0.1 M $H_2SO_4$, showing well-defined peaks for the surface oxidation of single crystal domains as early as in the fifth cycle.[44,45] These peaks are thought to arise from the removal of Ag impurities that relax the high energy low-coordinated atomic sites on the surface.[44] Our NPGF powder, however, shows very stable CVs after the first cycle, implying that it has stable energetic atomic sites on the surface. In addition, our NPGF powder shows only about 16 % of ECSA loss (Supplementary Fig. 12a) which is small considering the number of cycles.[44,46] The 100$^{th}$ potential cycle in 0.1 M KOH solution (Fig. 4b) demonstrates that surface oxidation and reduction take place in the range of 0.6 V ~ 1.6 V. The pre-oxidation peak at potentials below 1.2 V has also been observed in conventional Au and is related to the chemisorption of OH$^-$ anions onto the Au surface. The main peaks during the positively going scan appear at 1.25, 1.4, and 1.5 V while the peaks forming during the negatively going scan appear at 1.1 V, which agrees very well with previously reported measurements on polycrystalline Au.[43] After cycling 200 times in alkaline media, the final ECSA loss is around 32 % (Supplementary Fig. 12b). As can be observed from the CV, NPGF oxidizes more easily in alkaline solution than in acidic media, starting the oxidation in KOH solution 100 mV lower than in $H_2SO_4$ solution. This early oxidation makes the material more susceptible to coarsening upon prolonged cycling in this electrolyte. Nevertheless, the ECSA reduction is still a modest change after 200 cycles (Supplementary Fig. 13). As explained by the TEM images, this is attributed to the high density of grain boundaries, which increase the energy barrier for metal atom diffusion compared to the surface diffusion in a dealloyed NPM structure. Our NPGFs are particularly promising for (electro)catalysis, as they possess many grain boundaries and multiple surface directions that are nevertheless structurally robust.

----- FIGURE 4 ------

Under potential deposition (UPD) of lead (Pb) was carried out to examine the change of surface structures of a NPGF powder "as-prepared" (Supplementary Fig. 14a), "pre-cycled in acid" solution (Supplementary Fig. 14b), and "pre-cycled in alkaline" solution (Supplementary Fig. 14c). The CV clearly shows well-resolved UPD signals. Two intense peaks in the negatively going scan appear at 0.35 to 0.40V and 0.47 to 0.52 V and correspond to the Pb UPD on Au {111} and {110} domains, respectively. Two intense peaks in the positive going scan correspond to the anodic peaks at 0.45 to 0.50 V and 0.60 to 0.65 V attributed to oxidation of the Pb UPD layer. There is an additional peak between 0.35 to 0.40 V that has been reported to originate from defects near {111} domains. Note, that all peaks are very broad compared to the results from a single crystal Au electrode,[47] indicating that each domain is most likely surrounded by defects. This result illustrates that the NPGF contains many low-coordinated atoms, such as steps and kinks. There is no significant change in the peak current and potential even after cycling in an acidic solution, which is a remarkable difference to the reported UPD results of dealloyed nanoporous Au.[48] This observation demonstrates again that the surface structure with high surface energy in the nanoporous gold films is more robust than that of dealloyed nanoporous Au even under operating condition,[48] making the new material ideal for (electro)catalysis.

----- FIGURE 5 ------

The electrocatalytic performance of the NPGF powder was confirmed investigating the electrochemical methanol oxidation. Figure 5a shows the CV of the NPGF powder in 0.1 M KOH with and without 1 M methanol. The NPGF powder was pre-cycled in KOH solution before methanol oxidation. Single-crystal Au (111) is a less active catalyst for methanol oxidation, especially in an acidic solution. However, Au with nanoscale roughness shows a higher activity due to the presence of low-coordinated atoms and surface defects.[47,49–51] Our NPGF powder clearly facilitates strong methanol oxidation. Especially the shape of the CV curve during methanol oxidation is very similar to that of dealloyed nanoporous Au *after* it has been pre-cycled in acid (to dissolve Ag impurities).[44] This means that our NPGF is pure and exhibits a high surface area with many nanoscale pores and ligaments containing ideal surface structures for catalytic reactions. HR TEM was measured for the sample after being used as electrocatalyst for methanol oxidation to compare the atomic structure and the surface facets before and after the electrochemical reactions. As can be seen in Supplementary Fig. 15, the NPGF powder still has many facets in the curved region. Also, CVs of NPGF powder after MOR at constant potential for 2 h show negligible change, together with only 2 % loss of ECSA (Supplementary Fig. 16). These results reveal that the structure is not affected during the MOR and remains highly catalytic.

To evaluate other nanoporous metals and their application as electrocatalysts, a nanoporous nickel film (NPNF) was also fabricated and examined as a catalyst for the OER. Figure 5b shows the CV of NPNF powder performed in 0.1 M NaOH over 10 cycles, where the Ni oxidation peak appears at ~1.4 V (in the forward scan) and the reduction peak appears at ~1.3 V (in the reverse scan). Furthermore, linear sweep voltammetry (LSV) was also performed on the NPNF (shown in Supplementary Fig. 17a), in the same electrolyte. We note that the current density obtained for the NPNF are very high compared to previous reports[52–55] at an overpotential of circa 0.4 V (which is here the potential versus the potential for oxygen evolution "OER", 1.23 V vs. RHE), reaching more than 100 mA/cm$^2$ (the geometric area of the electrode was used, see experimental section). The onset potential is determined to be 284 mV for the NPNF powder, which is extremely low for a Ni-based OER electrocatalyst.[52–55] We associate the low onset potential measured for the OER on the NPNF powder with the high porosity, curved surfaces, and many grain boundaries as shown in the TEM images in the inset of Fig. 5b. Multiple crystal directions are observed in curved regions (Supplementary Fig. 18), which all can form highly active sites reducing the activation energy for the OER to occur. In addition, the sheet resistance was measured to be around 162.1 Ω/sq, which clearly shows that the NPNF powder is well-connected and electrically conductive contributing to high-performance OER. Moreover, we also calculated the Tafel slope of the NPNF (Supplementary Fig. 17b) and obtained a slope of 76 mV/dec, which is lower than in previous reports,[56] further indicating the exceptional performance of the NPNF powder for OER. These results show that the NPMFs can be fabricated with different metals and that these are promising for several applications, especially as electrocatalysts that benefit from the very high activity.

## 3. Conclusions

Plasma-based dry synthesis is a facile and general method to produce NPMFs. The majority of methods have relied on dealloying that cannot avoid the presence of the less noble metal (or sacrificial metal) in the obtained nanoporous structure. Our method fully overcomes the requirement of a master alloy and a wasteful dissolution process (typical master alloys contain more than 60 % of the sacrificial metal) and instead produces a pristine material, free of residues of additional compounds. Other methods including electrochemical deposition, template-assisted methods, and sol-gel synthesis do not achieve the curved structures at the nanoscale that our method provides. Apart from these, vapor dealloying[57] and phase boundary gelation[9] were reported to make 2D metal networks, but they are not as general as our method, *i.e.*, they only work for a particular metal and are not suitable for large-area fabrication. Since our plasma-based dry synthesis does not include wet-chemical reaction steps, it can be applied to a large number of metals including Au, Ag, Pt, Pd, Ni, Fe, and even alloys. Considering that physical vapor deposition is used in the process, any metals that can be evaporated can be used in our scheme to form NPMFs. This opens up the opportunity to employ co-deposition and hence extend the fabrication to nanoporous alloy metal films. The plasma condition can be adjusted to the nature of the metal film. Non-noble metal films were obtained using a mixture of Ar and $H_2$ to prevent the oxidation of the metal during the removal of the PMMA film.

Compared to conventional Au NPs, our NPGF has the clear advantage that it is mechanically more rigid, whereas Au NPs may easily aggregate when the experimental conditions change the colloidal stability. The NPGF is also connected and thus electrically conducting, which is a major advantage for electrochemical applications. Moreover, our NPGF is catalytically highly active due to many curved structures that have high surface energy, whereas the surfaces of Au NPs often have energetically stable facets. Even compared with conventional dealloyed nanoporous Au, our films are very stable, as revealed by potential cycling in acids and bases and Pb underpotential deposition. Our NPGF fabricated by dry synthesis can further be used to gain an understanding of the methanol oxidation mechanism in nanoporous Au, as our scheme produces highly stable NPGF of high purity and controllable morphology.

The method described herein yields ultra-thin NPM structures, that can be tuned in material composition and pore structure and thus suggest further promising applications as transparent conducting electrodes (see Supplementary Fig. 19, Supplementary Table 2). It is also possible to lift off the films from the substrate and realize free-standing NPMFs (Supplementary Fig. 20). The NPGF was fabricated in our current setup on a 3-inch wafer as shown in Supplementary Fig. 21, showing very uniform and similar morphologies across the entire area. However, our dry synthesis method is not limited to a specific size, and can potentially be adapted to the large-area fabrication of NPMFs. Furthermore, our scheme can be extended to produce 3D NPM structures by stacking and folding the NPMFs. Supplementary Fig. 22 shows a stacked nanoporous Au structure with a thickness approaching 1 µm. We expect that our dry synthesis method, which is robust and which can reproducibly yield impurity-free films, will open up new opportunities in the application of NPMs.

## 4. Experimental Section

*Fabrication of NPMs:* Poly(methyl methacrylate) (PMMA, average Mw ~ 120,000 g/mol from Sigma Aldrich) was dissolved (1.5 wt. %) in chloroform for 12 h at room temperature using magnetic stirring. The solution was spin-coated on a Si wafer substrate (boron-doped, <100> orientation, native oxide covered, cleaned in Piranha solution (a mixture of 3 volume parts of concentrated $H_2SO_4$ and 1 volume part of 30% $H_2O_2$ for 30 min)) at 1000 rpm for 1 min. Each metal (> 99.99 % purity) was evaporated by e-beam on a PMMA thin film at room temperature with an oblique angle of 80 °, with a rate of 0.05 nm/s, 10 nm target thickness considering tooling factor, and rotation speed 0.72 °/s. The deposited Au film was plasma-treated in 0.4 mbar air ambient with 200 W for 15 min. The Ag NPMF was obtained after treatment in 0.4 mbar Ar ambient with 200 W for 15 min, the Ni NPMF in 0.4 mbar W10 ambient (Ar 90 %, $H_2$ 10 %) with 150 W for 15 min, and other NPMFs (Pt, Pd, Fe) in 0.4 mbar W10 ambient with 300 W for 15 min.

*Compositional analysis:* The nanoporous gold film was fabricated on a Si substrate. After complete etching of PMMA, the film was used for combustion analysis and XPS. A PMMA film coated on Si substrate without a metal layer was also characterized, for reference. For the combustion analysis, a sample was burned in a flowing stream of oxygen.[35] Very low concentrations of carbon can be detected (resolution down to 1 ppm (wt.)). We compared the absorption spectra of three NPGFs on Si substrates and three cleaned bare Si substrates. Any $CO_2$ was detected by absorption of infrared radiation. The measurement was repeated three times. XPS measurements were performed on a Theta Probe Angle-Resolved X-ray Photoelectron Spectrometer System (Thermo Fisher Scientific Inc.). The base pressure was $3\times10^{-10}$ mbar and the excitation X-ray source was a monochromatic Al Kα radiation (100 W, $hv$ = 1486.68 eV). Survey spectra were measured at a pass energy of 200 eV, followed by high-resolution spectra for Au 4f,

C 1s, O 1s, and Si 2p with a pass energy of 10 eV, and a step size of 0.05 eV. Charge was corrected for all the binding energies by shifting the C-C (or C-H) part of the C 1s peak to 284.8 eV. Peak fitting was performed using the Avantage software (version 5.9904).

*Sample preparation for electrochemistry:* To remove the NPGF off the substrate, a thicker PMMA film (5 wt. % spin-coated at 1000 rpm for 1 min) was used. After plasma treatment, the PMMA film was not entirely removed and was thus able to support the NPGF. This sample was immersed in pure acetone to dissolve the PMMA film to obtain a free-standing NPGF that could be lifted off. NPGF was collected in the Teflon Petri dish and dried for 3 h in Ar at ambient conditions. The chunks of the film obtained by this method were the starting material to be loaded into CMEs for the electrochemical experiments. CMEs were produced by sealing a 1.5 cm Au wire of 100 µm diameter (99.99+%, Goodfellow, Friedberg, Germany) in a borosilicate glass capillary which was previously heated to form a tip. The assembly was ground with grade 1500 abrasive paper to expose the disk-shape cross-section of the Au wire followed by polishing with abrasive papers of 15, 9, 3, and 1 µm grain size. Subsequently, the disk-shape cross section is polished to mirror finish using a microgrinder (EG-401, Narishige, Tokyo, Japan). Finally, the Au wire inside the capillary was connected to a Cu wire with silver-epoxy glue (EPO-TEKs, John P. Kummer GmbH, Germany). The cavities were formed by potentiostatic dissolution of the exposed Au at +1.1 V vs. a saturated calomel electrode (SCE, from ALS, Tokyo, Japan) in 1 M HCl during 60 s and afterwards cycled 5 times at 100 mV s$^{-1}$ between 0.5 V and 1.5 V vs. SCE. The quality of the polish on the microelectrodes and the depth of the cavities [(21 ± 1) µm] were verified using a confocal laser scanning microscope (TCS SP2, Leica Microsystems GmbH, Germany) with a HC PL Fluotar 50×/0.8 dry lens. The cavity is cleaned by immersion in Piranha solution for 1 h and transferred to ultrapure water for at least 24 h. The cavity was filled by slightly pressing the CME into the NPGF powder that has been lifted off (as described above). The end of the filled CME was washed with ultrapure water and any excess NPGF outside the cavity was wiped off with a soft cloth. The filling is monitored by inspection with a 6× magnifying lens. The electrochemical cell is built with a 25 mL glass vial and a Teflon cap with five necks. A CME filled with the sample was used as the working electrode and an Au coil as the auxiliary electrode.

*Characterization of electrochemical properties:* The reference electrodes were Hg/HgSO$_4$/K$_2$SO$_4$(Sat) (from ALS, Japan) in Cl$^-$-free acid media and a Hg/HgO/1 M NaOH (from

ALS, Japan) in alkaline media. Electrochemical characterization was performed on a PGSTAT128N potentiostat with an analog scan generator (Autolab-Metrohm, Filderstadt, Germany with scan250 and NOVA 2.1 software). All glassware was cleaned by immersion in 1 g/L KMnO$_4$ solution acidified with 20 mL/L 96 % H$_2$SO$_4$ for at least 24 h. To remove the excess of MnO$_4^-$, the glassware was immersed in 40 mL/L 30 % H$_2$O$_2$ solution acidified with 20 mL/L of 96 % H$_2$SO$_4$, until no visual evidence of purple color was detected. Finally, it was rinsed with ultrapure water and boiled three times. For the UPD experiments, an alkaline Pb-containing solution was prepared by dissolving the weighted amount of KOH in half of the final volume, followed by the addition of Pb(NO$_3$)$_2$ and finally filling up water to the final volume. This order is important to avoid the oxidation of Pb$^{2+}$ to Pb$^{4+}$ promoted by the dissolved oxygen in strongly alkaline medium. All potentials are reported versus the reversible hydrogen electrode (RHE) by using the equation $E_{RHE} = E_{SHE} + 0.140$ V $+ (0.059 \cdot pH)$, where the potential against the standard hydrogen electrodes is obtained by constant potential distance to the used laboratory reference electrodes $E_{SHE} = E_{Hg|HgO|1\ M\ NaOH} + 0.14$ V or $E_{SHE} = E_{Hg|HgSO4|K2SO4(sat)} + 0.64$ V. The OER measurements of the NPNF powder were performed in a setup similar to the one used for the MOR of NPGF powder. The working electrode was a CME filled with NPNF powder. The counter electrode was a high-surface area Pt mesh electrode and the reference electrode was a Ag|AgCl| KCl 3 M ($E° = 195$ mV). The measurements were performed in a standard three-electrode cell system using an Autolab potentiostat (PGSTAT204, Metrohm) in conjunction with the Nova software (2.1, Metrohm). All measurements were performed in 0.1 M aqueous (MilliQ) NaOH solution (99.99% Suprapur, Merck), which was bubbled with Ar for 2 h prior to measurements. The pH of the solution was 12.6 and a gentle flow of Ar was kept over the solution throughout the experiments to create an Ar mantle above the solution and prevent oxygen diffusion into the solution. The CV measurements were performed with a scan rate of 10 mV s$^{-1}$, with a potential step of 2.44 mV for 10 cycles, followed by the LSV measurements, which were performed with a scan rate of 1 mV s$^{-1}$, and a potential step of 0.5 mV. The WE current was normalized to the microelectrode geometric disc area, taken as 7.8·10$^{-5}$ cm$^2$. The potentials were corrected to the RHE scale by the following equation:

$E_{RHE} = E_{Ag|AgCl} + (0.059*pH) + E°_{Ag|AgCl}$

The OER overpotentials were calculated by subtracting the thermodynamic potential of the OER, $E° = 1.23$ V, from the measured RHE-corrected potentials.

*Electron microscopy:* For SEM, a NPGF was used that has been fabricated by the method described herein on a Si substrate after complete etching of PMMA. For the TEM preparation of the "as-prepared" NPGF, the same method was used as for the preparation before the electrochemical measurement. The PMMA film was not etched entirely, and the sample was immersed in pure acetone to dissolve the PMMA film to obtain a free-standing Au film that could be lifted-off. (Supplementary Fig. 23) The film was then picked up with a plasma-cleaned TEM carbon grid, followed by drying in Ar flow for an hour. For TEM-EDX measurement, a $SiO_2$ grid was used to avoid strong carbon signals from the grid. For HR TEM of the NPGF after methanol oxidation, the Au-film filled microelectrode was sonicated in pure ethanol (50 μL) for 3 min in an Eppendorf tube. Then 5 μL of the solution was drop-casted on a plasma-cleaned TEM carbon grid, and dried in Ar flow for 5 min. Solution casting and drying was repeated 10 times. A 200 kV JEOL ARM200CF scanning transmission electron microscope equipped with a cold field emission electron source, and a CETCOR image corrector (CEOS GmbH) was used to obtain HR TEM images, and the ZEISS SESAM (sub-electronvolt-sub-angstrom-microscope) with a field emission gun (200 kV), equipped with a monochromator, MANDOLINE-filter and a 60 $mm^2$ Thermo Fischer ultradry EDX detector was used to obtain the TEM EDX data.

**Supporting Information**

Supporting Information is available from the Wiley Online Library or from the author.

**Conflict of Interest**

The authors declare no conflict of interest.

**Data availability**

**References**

1. Tappan, B. C., Steiner III, S. A. & Luther, E. P. Nanoporous metal foams. *Angew. Chem. Int. Ed.* **49**, 4544–4565 (2010). DOI: 10.1002/anie.200902994

2. Koya, A. N., Zhu, X., Ohannesian, N., Yanik, A. A., Zaccaria, R. P., Krahne, R., Shih, W.–C. & Garoli, D. Nanoporous metals: from plasmonic properties to applications in enhanced spectroscopy and photocatalysis. *ACS Nano* **15**, 6038–6060 (2021) DOI: 10.1021/acsnano.0c10945

3. Zhang, R. & Olin, H. Porous gold films—a short review on recent progress. *Materials* **7**, 3834–3854 (2014). DOI: 10.3390/ma7053834

4. Jin, T., Terada, M., Bao, M. & Yamamoto, Y. Catalytic performance of nanoporous metal skeleton catalysts for molecular transformations. *ChemSusChem* **12**, 2936–2954 (2019). DOI: 10.1002/cssc.201900318

5. Khristosov, M. K., Dishon, S., Noi, I., Katsman, A. & Pokroy, B. Pore and ligament size control, thermal stability and mechanical properties of nanoporous single crystals of gold. *Nanoscale* **9**, 14458–14466 (2017). DOI: 10.1039/C7NR04004K

6. Fujita, T., Guan, P., McKenna, K., Lang, X., Hirata, A., Zhang, L., Tokunaga, T., Arai, S., Yamamoto, Y., Tanaka, N., Ishikawa, Y., Asao, N., Yamamoto, Y., Erlebacher, J. & Chen, M. Atomic origins of the high catalytic activity of nanoporous gold. *Nat. Mater.* **11**, 775–780 (2012). DOI: 10.1038/nmat3391

7. Fajín, J. L. C., Cordeiro, M. N. D. S. & Gomes, J. R. B. On the theoretical understanding of the unexpected $O_2$ activation by nanoporous gold. *Chem. Commun.* **47**, 8403–8405 (2011). DOI: 10.1039/C1CC12166A

8. Wang, H., Fang, Q., Gu, W., Du, D., Lin, Y. & Zhu, C. Noble metal aerogels. *ACS Appl. Mater. Interfaces* **12**, 52234–52250 (2020). DOI: 10.1021/acsami.0c14007

9. Hiekel, K., Jungblut, S., Georgi, M. & Eychmüller, A. Tailoring the morphology and fractal dimension of 2D mesh-like gold gels. *Angew. Chem. Int. Ed.* **132**, 12146–12152 (2020). DOI: 10.1002/ange.202002951

10. Welch, A. J., DuChene, J. S., Tagliabue, G., Davoyan, A., Cheng, W.–H. & Atwater, H. A. Nanoporous gold as a highly selective and active carbon dioxide reduction catalyst. *ACS Appl. Energy Mater.* **2**, 164–170 (2019). DOI: 10.1021/acsaem.8b01570

11. Chen, Q., Ding, Y. & Chen, M. Nanoporous metal by dealloying for electrochemical energy conversion and storage. *MRS Bull.* **43**, 43–48 (2018). DOI: 10.1557/mrs.2017.300

12. Qiu, H.–J., Li, X., Xu, H.–T., Zhang, H.–J. & Wang, Y. Nanoporous metal as a platform for electrochemical and optical sensing. *J. Mater. Chem. C* **2**, 9788–9799 (2014). DOI: 10.1039/C4TC01913J

13. Vidal, C., Wang, D., Schaaf, P., Hrelescu, C. & Klar, T. A. Optical plasmons of individual gold nanosponges. *ACS Photonics* **2**, 1436–1442 (2015). DOI: 10.1021/acsphotonics.5b00281

14. Erlebacher, J., Aziz, M. J., Karma, A., Dimitrov, N. & Sieradzki, K. Evolution of nanoporosity in dealloying. *Nature* **410**, 450–453 (2001). DOI: 10.1038/35068529

15. Pedireddy, S., Lee, H. K., Tjiu, W. W., Phang, I. Y., Tan, H. R., Chua, S. Q., Troadec, C. & Ling, X. Y. One-step synthesis of zero-dimensional hollow nanoporous gold nanoparticles with enhanced methanol electrooxidation performance. *Nat. Commun.* **5**, 4947 (2014). DOI: 10.1038/ncomms5947

16. Khristosov, M. K., Bloch, L., Burghammer, M., Kauffmann, Y., Katsman, A. & Pokroy, B. Sponge-like nanoporous single crystals of gold. *Nat. Commun.* **6**, 8841 (2015). DOI: 10.1038/ncomms9841

17. Lee, D. H., Park J. S., Hwang J. H., Kang, D. H., Yim, S.–Y. & Kim, J. H. Fabrication of hollow nanoporous gold nanoshells with high structural tunability based on the plasma etching of polymer colloid templates. *J. Mater. Chem. C* **6**, 6194-6199 (2018). DOI: 10.1039/C8TC01658E

18. Rebbecchi, T. A. & Chen, Y. Template-based fabrication of nanoporous metals. *J. Mater. Res.* **33**, 2–15 (2018). DOI: 10.1557/jmr.2017.383

19. Haupt, M., Miller, S., Glass, R., Arnold, M., Sauer, R., Thonke, K., Möller, M. & Spatz, J. P. Nanoporous gold films created using templates formed from self-assembled structures of inorganic–block copolymer micelles. *Adv. Mater.* **15**, 829–831 (2003). DOI: 10.1002/adma.200304688

20. Bartlett, P. N., Baumberg, J. J., Birkin, P. R., Ghanem, M. A. & Netti, M. C. Highly ordered macroporous gold and platinum films formed by electrochemical deposition through templates assembled from submicron diameter monodisperse polystyrene spheres. *Chem. Mater.* **14**, 2199–2208 (2002). DOI: 10.1021/cm011272j

21. Attard, G. S., Bartlett, P. N., Coleman, N. R. B., Elliott, J. M., Owen, J. R., & Wang, J. H. Mesoporous platinum films from lyotropic liquid crystalline phases. *Science* **278**, 838–840 (1997) DOI: 10.1126/science.278.5339.838

22. Qian, F., Troksa, A., Fears, T. M., Nielsen, M. H., Nelson, A. J., Baumann, T. F., Kucheyev, S. O., Han, T. Y.–J., & Bagge–Hansen, M. Gold aerogel monoliths with tunable ultralow densities. *Nano Lett.* **20**, 131–135 (2020). DOI: 10.1021/acs.nanolett.9b03445

23. Liu, P., Guan, P., Hirata, A., Zhang, L., Chen, L., Yuren, W., Ding, Y., Fujita, T., Erlebacher, J. & Chen, M. Visualizing under-coordinated surface atoms on 3D nanoporous gold catalysts. *Adv. Mater.* **28**, 1753–1759 (2016). DOI: 10.1002/adma.201504032

24. Guo, X., Zhang, C., Tian, Q. & Yu, D. Liquid metals dealloying as a general approach for the selective extraction of metals and the fabrication of nanoporous metals: a review. *Mater. Today Commun.* **26**, 102007 (2021). DOI: 10.1016/j.mtcomm.2020.102007

25. McCue, I., Benn, E., Gaskey, B. & Erlebacher, J. Dealloying and dealloyed materials. *Annu. Rev. Mater. Res.* **46**, 263–286 (2016). DOI: 10.1146/annurev-matsci-070115-031739

26. Graf, M., Haensch, M., Carstens, J., Wittstock, G. & Weissmüller, J. Electrocatalytic methanol oxidation with nanoporous gold: microstructure and selectivity. *Nanoscale* **9**, 17839–17848 (2017). DOI: 10.1039/C7NR05124G

27. Lackmann, A., Bäumer, M., Wittstock, G. & Wittstock, A. Independent control over residual silver content of nanoporous gold by galvanodynamically controlled dealloying. *Nanoscale* **10**, 17166–17173 (2018). DOI: 10.1039/C8NR03699C

28. Wittstock, A., Neumann, B., Schaefer, A., Dumbuya, K., Kübel, C., Biener, M. M., Zielasek, V., Steinrück, H.–P., Gottfried, J. M., Biener, J., Hamza, A. & Bäumer, M. Nanoporous Au: an unsupported pure gold catalyst? *J. Phys. Chem. C* **113**, 5593–5600 (2009). DOI: 10.1021/jp808185v

29. Jin, Y., Li, R. & Zhang, T. Formation of nanoporous silver by dealloying Ca–Ag metallic glasses in water. *Intermetallics* **67**, 166–170 (2015). DOI: 10.1016/j.intermet.2015.08.011

30. Madern, N., Monnier, J., Cachet–Vivier, C., Zhang, J., Bastide, S., Paul–Boncour, V. & Latroche, M. Anisotropic nanoporous nickel obtained through the chemical dealloying of $Y_2Ni_7$ for the comprehension of anode surface chemistry of Ni-*M*H batteries. *ChemElectroChem* **6**, 5022–5031 (2019). DOI: 10.1002/celc.201901281

31. Pashley, D. W., Stowell, M. J., Jacobs, M. H. & Law, T. J. The growth and structure of gold and silver deposits formed by evaporation inside an electron microscope. *Philos. Mag. J. Theor. Exp. Appl. Phys.* **10**, 127–158 (1964). DOI: 10.1080/14786436408224212

32. Winkler, K., Wojciechowski, T., Liszewska, M., Górecka, E. & Fiałkowski, M. Morphological changes of gold nanoparticles due to adsorption onto silicon substrate and oxygen plasma treatment. *RSC Adv.* **4**, 12729–12736 (2014). DOI: 10.1039/C4RA00507D

33. Hoffman, R. W. Stresses in thin films: the relevance of grain boundaries and impurities. *Thin Solid Films* **34**, 185–190 (1976). DOI: 10.1016/0040-6090(76)90453-3

34. Jeffers, G., Dubson, M. A. & Duxbury, P. M. Island-to-percolation transition during growth of metal films. *J. Appl. Phys.* **75**, 5016–5020 (1994). DOI: 10.1063/1.355742

35. López, G. A. & Mittemeijer, E. J. The solubility of C in solid Cu. *Scr. Mater.* **51**, 1–5 (2004). DOI: 10.1016/j.scriptamat.2004.03.028

36. Moulder, J. F. Handbook of X-ray photoelectron spectroscopy: a reference book of standard spectra for identification and interpretation of XPS data. Physical Electronics Division, Perkin-Elmer Corporation, (1992).

37. Passiu, C., Rossi, A., Weinert, M., Tysoe, W., & Spencer N. D. Probing the outermost layer oft hin gold films by XPS and density functional theory. *Appl. Surf. Sci.* **507**, 145084 (2020). DOI: 10.1016/j.apsusc.2019.145084

38. Yamamoto, M., Matsumae, T., Kurashima, Y., Takagi, H., Suga, T., Itoh, T. & Higurashi, E. Comparison of argon and oxygen plasma treatments for ambient room-temperature wafer-scale Au–Au bonding using ultrathin Au films. *Micromachines* 10, 119 (2019). DOI: 10.3390/mi10020119

39. Piao, H., Fairley, N. & Walton, J. Application of XPS imaging analysis in understanding interfacial delamination and X-ray radiation degradation of PMMA. *Surf. Interface Anal.* **45**, 1742–1750 (2013). DOI: 10.1002/sia.5316

40. Mullins, W. W. Two-dimensional motion of idealized grain boundaries. *J. Appl. Phys.* **27**, 900–904 (1956). DOI: 10.1063/1.1722511

41. Upmanyu, M., Smith, R. W. & Srolovitz, D. J. Atomistic simulation of curvature driven grain boundary migration. *Interface Science* **6**, 41–58 (1998). DOI: 10.1023/A:1008608418845

42. Winterbottom, W. L. Equilibrium shape of a small particle in contact with a foreign substrate. *Acta Metall.* **15**, 303–310 (1967). DOI: 10.1016/0001-6160(67)90206-4

43. Zhang, J., Liu, P., Ma, H. & Ding, Y. Nanostructured porous gold for methanol electro-oxidation. *J. Phys. Chem. C* **111**, 10382–10388 (2007). DOI: 10.1021/jp072333p

44. Silva Olaya, A. R., Zandersons, B. & Wittstock, G. Restructuring of nanoporous gold surfaces during electrochemical cycling in acidic and alkaline media. *ChemElectroChem* **7**, 3670–3678 (2020). DOI: 10.1002/celc.202000923

45. Wang, Z., Nign, S., Liu, P., Ding, Y., Hirata, A., Fujita, T. & Chen, M. Tuning surface structure of 3D nanoporous gold by surfactant-free electrochemical potential cycling. *Adv. Mater.* **29**, 1703601 (2017). DOI: 10.1002/adma.201703601

46. Rurainsky, C., Manjón, A. G., Hiege, F., Chen, Y. –T., Scheu, C. & Tschulik, K. Electrochemical dealloying as a tool to tune the porosity, composition and catalytic activity of nanomaterials. *J. Mater. Chem. A* **8**, 19405–19413 (2020). DOI: 10.1039/D0TA04880A

47. Hernández, J., Solla-Gullón, J., Herrero, E., Aldaz, A. & Feliu, J. M. Methanol oxidation on gold nanoparticles in alkaline media: unusual electrocatalytic activity. *Electrochim. Acta* **52**, 1662–1669 (2006). DOI: 10.1016/j.electacta.2006.03.091

48. Silva Olaya, A. R., Zandersons, B. & Wittstock, G. Effect of the residual silver and adsorbed lead anions towards the electrocatalytic methanol oxidation on nanoporous gold in alkaline media. *Electrochim. Acta* **383**, 138348 (2021). DOI: 10.1016/j.electacta.2021.138348

49. Rodriguez, P. & Koper, M. T. M. Electrocatalysis on gold. *Phys. Chem. Chem. Phys.* **16**, 13583–13594 (2014). DOI: 10.1039/C4CP00394B

50. Borkowska, Z., Tymosiak-Zielinska, A. & Shul, G. Electrooxidation of methanol on polycrystalline and single crystal gold electrodes. *Electrochim. Acta* **49**, 1209–1220 (2004). DOI: 10.1016/j.electacta.2003.09.046

51. Borkowska, Z., Tymosiak-Zielinska, A. & Nowakowski, R. High catalytic activity of chemically activated gold electrodes towards electro-oxidation of methanol. *Electrochim. Acta* **49**, 2613–2621 (2004). DOI: 10.1016/j.electacta.2004.01.030

52. Roger, I., Shipman, M. A. & Symes, M. D. Earth-abundant catalysts for electrochemical and photoelectrochemical water splitting. *Nat. Rev. Chem.* **1**, 1–13 (2017). DOI: 10.1038/s41570-016-0003

53. Babar, N. –U. –A. & Joya, K. S. Spray-coated thin-film Ni-oxide nanoflakes as single electrocatalysts for oxygen evolution and hydrogen generation from water splitting. *ACS Omega* **5**, 10641–10650 (2020). DOI: 10.1021/acsomega.9b02960

54. Ali Akbari, M. S., Bagheri, R., Song, Z. & Najafpour, M. M. Oxygen-evolution reaction by nickel/nickel oxide interface in the presence of ferrate(VI) *Sci. Rep.* **10**, 8757 (2020). DOI: 10.1038/s41598-020-65674-x

55. McCrory, C. C. L., Jung, S., Peters, J. C. & Jaramillo, T. F. Benchmarking heterogeneous electrocatalysts for the oxygen evolution reaction. *J. Am. Chem. Soc.* **135**, 16977–16987 (2013). DOI: 10.1021/ja407115p

56. Yu, M., Moon, G., Bill, E. & Tüysüz, H. Optimizing Ni–Fe oxide electrocatalysts for oxygen evolution reaction by using hard templating as a toolbox. *ACS Appl. Energy Mater.* **2**, 1199–1209 (2019). DOI: 10.1021/acsaem.8b01769

57. Chauvin, A., Txia Cha Heu, W., Buh, J., Tessier, P.-Y. & El Mel, A.-A. Vapor dealloying of ultra-thin films: a promising concept for the fabrication of highly flexible transparent conductive metal nanomesh electrodes. *Npj Flex. Electron.* **3**, 5 (2019) DOI: 10.1038/s41528-019-0049-1

**Figures**

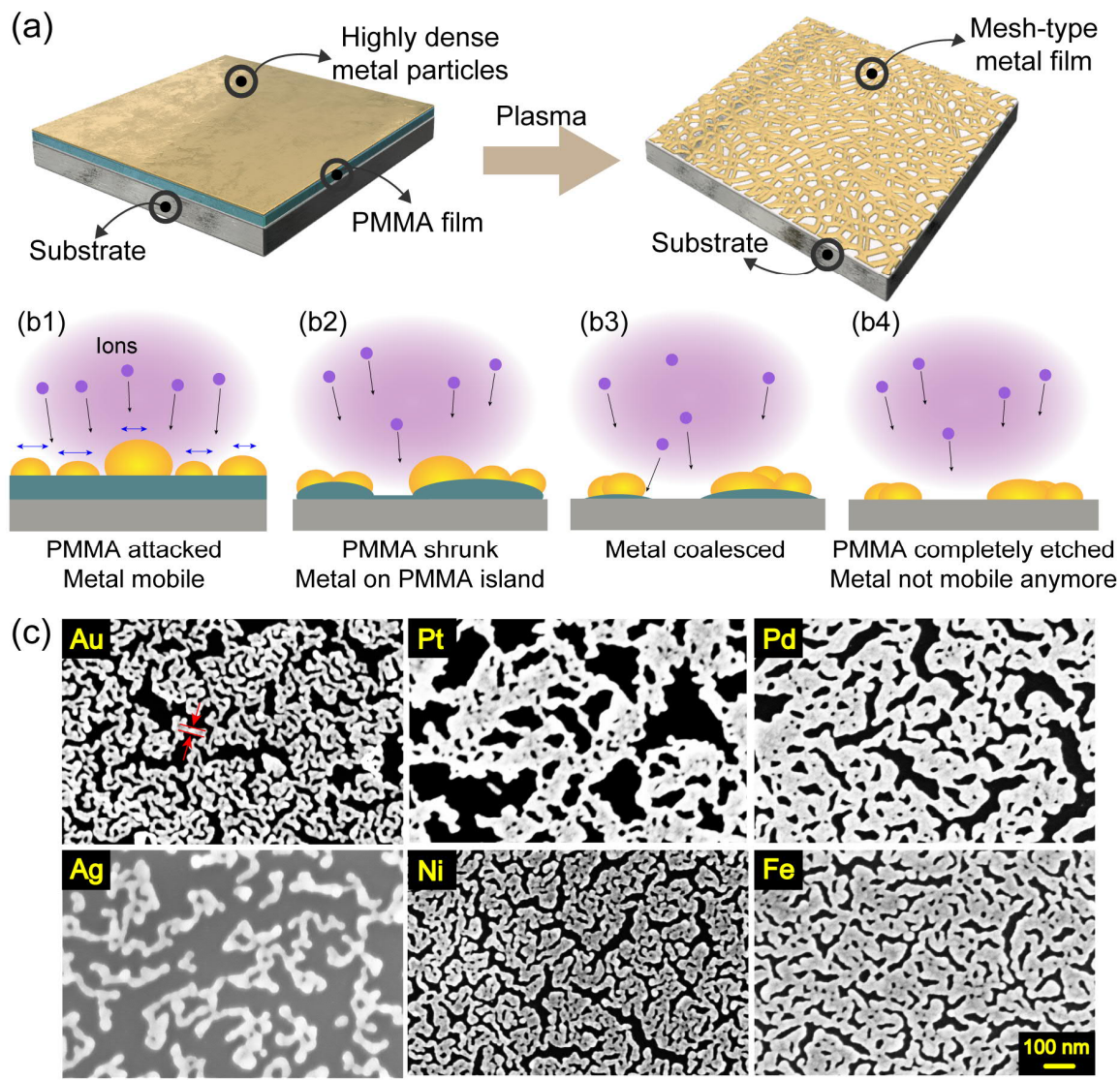

**Figure 1.** (a) Schematic: dry synthesis of nanoporous metals (NPMs). The deposited metal NPs on the PMMA layer transforms to NPM structures after plasma treatment. Cross-section schematics that show how the metal nanoparticles are thought to evolve over time (see text for details) are given in (b1) to (b4). (c) SEM images of different NPMs. The scale bar is 100 nm for all panels. Red lines with arrows indicate the ligaments of the structure.

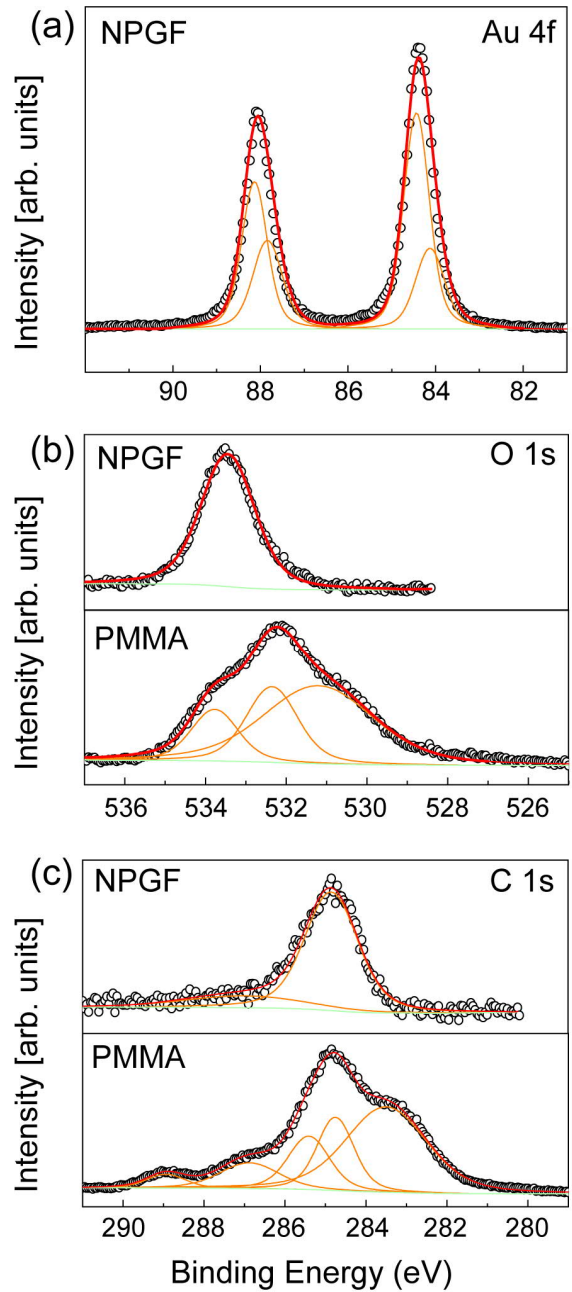

**Figure 2.** XPS results of NPGF sample. (a) Au 4f of NPGF sample. (b) Comparison of the O 1s spectrum and (c) C 1s spectrum from the NPGF sample and a spin-coated PMMA sample. Black dots are from the measurement and red lines are fitted curves.

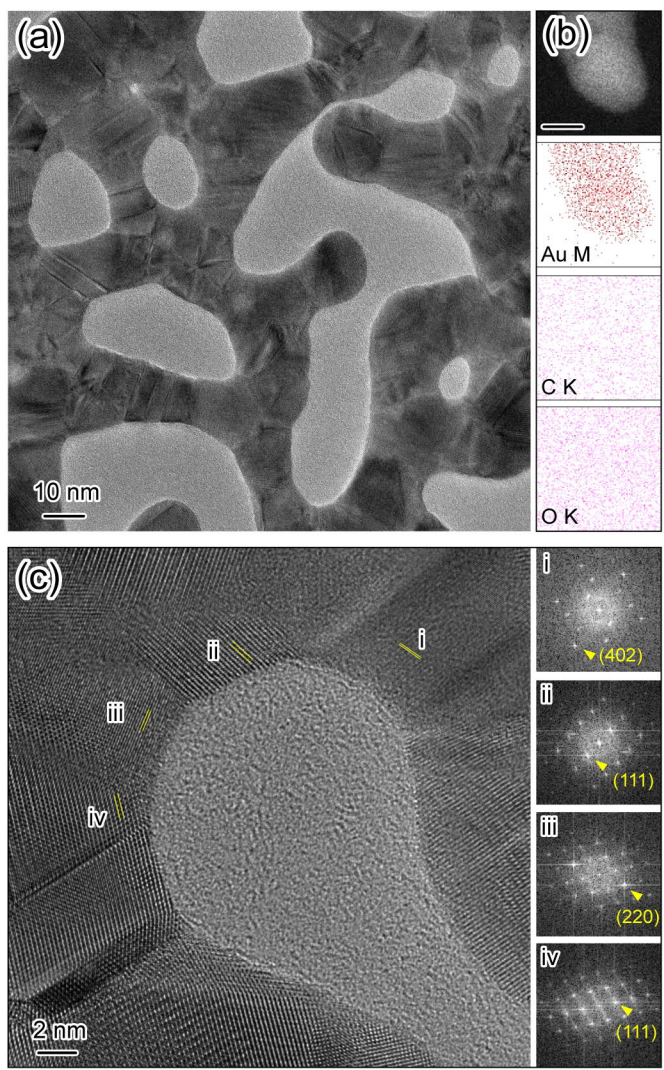

**Figure 3.** TEM measurements of NPGF. (a) Relatively low magnification image of NPGF showing highly curved ligaments and pores. Ligament sizes are 10 – 30 nm, which is relatively small compared to reported nanoporous Au structures made by dealloying. (b) TEM-EDX analysis of a specific protruded region of NPGF. Au M signal is clear, but C K and O K signals are not distinguishable from background signals. (c) Image of a concave region. FFT images were obtained from each region, and are shown in the right column in the order i to iv. Lattice spacings in yellow lines were measured from each grain. The NPGF has surfaces with multiple directions, especially in curved regions, and many grain boundaries between grains few nanometers in size.

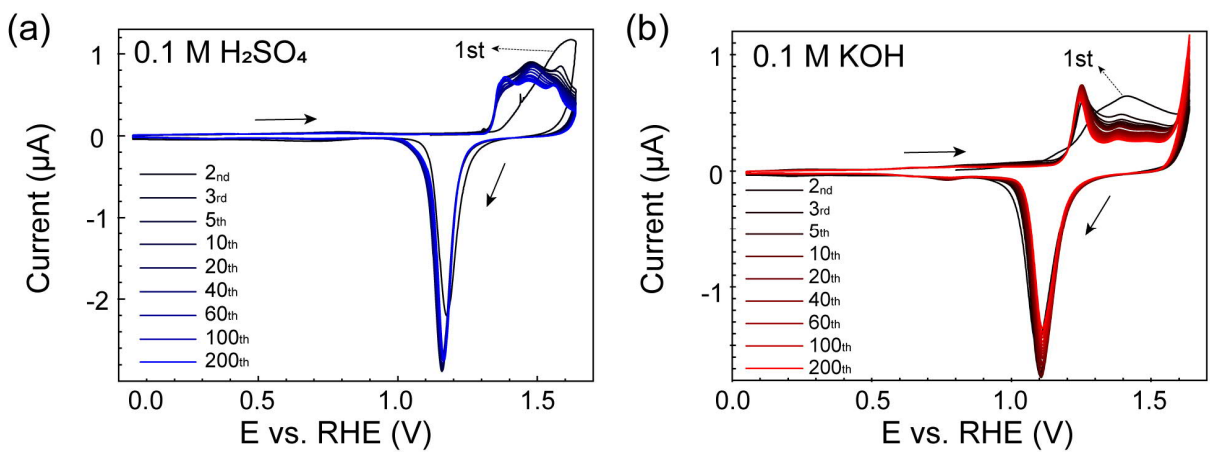

**Figure 4.** Voltammograms of NPGF powder in (a) 0.1 M H$_2$SO$_4$ (acidic, blue) and (b) 0.1 M KOH (alkaline, red) solutions over 200 cycles measured at 10 mV s$^{-1}$.

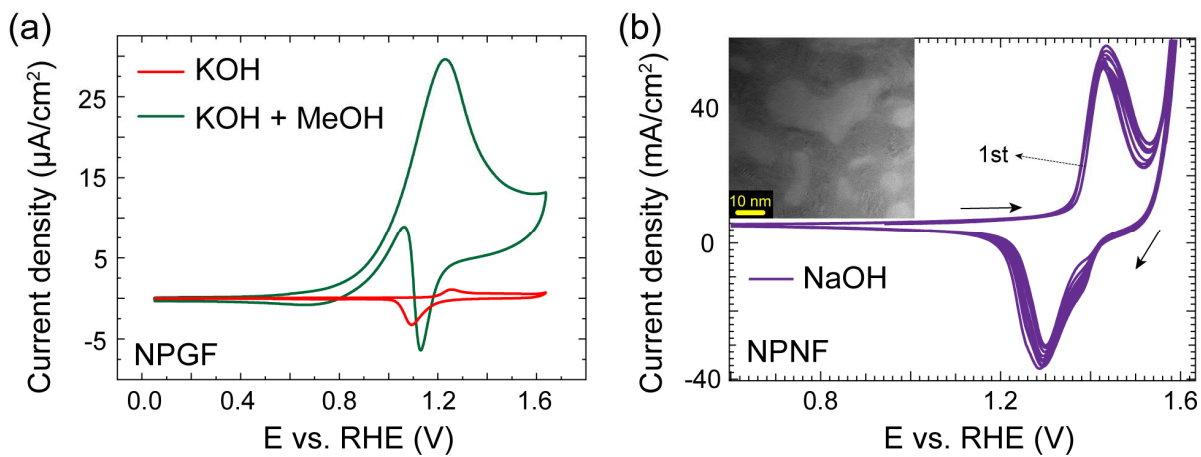

**Figure 5.** (a) Electro-oxidation of methanol at NPGF powder. Comparison of CVs in 0.1 M KOH solution and in 0.1 M KOH + 1 M methanol solution. Scan rate is 10 mV s$^{-1}$. The green curve shows strong methanol oxidation attributed to the reactive surface structures important for catalytic reactions. (b) Electrocatalytic analysis of NPNF powder. CV of 10 cycles in 0.1 M NaOH solution. Scan rate is 10 mV s$^{-1}$. The curve shows redox peaks for the formation of Ni oxides. Inset is a HR

TEM image of the NPNF powder *before* electrocatalytic measurement, showing many curved surfaces and grain boundaries.

**Supplementary Information**

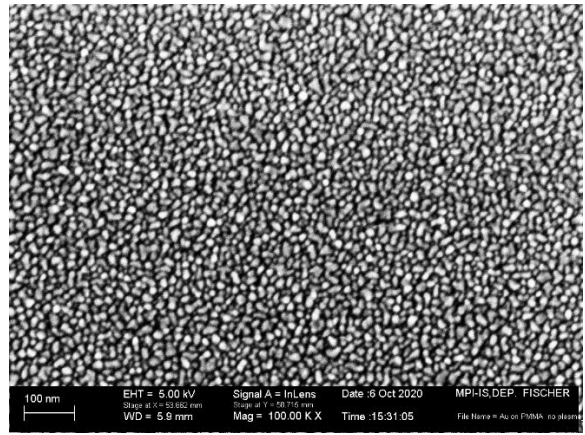

**Supplementary Figure 1**. Layer of Au nanoparticles on a PMMA substrate, deposited with oblique angle (or glancing angle deposition, GLAD) (50 nm, 80 deg).

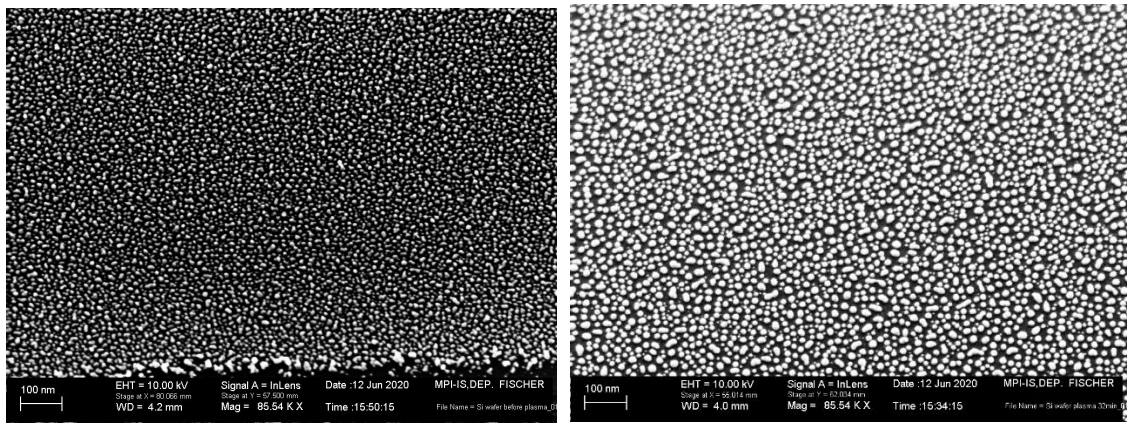

**Supplementary Figure 2**. Layer of Au nanoparticles on a Si wafer before (left) and after (right) plasma treatment. Several small (<10 nm) nanoparticles coalesce initially, but the resultant larger nanoparticles (larger than ~20 nm) are no longer mobile enough, and do not lead to networked structures on a Si wafer.

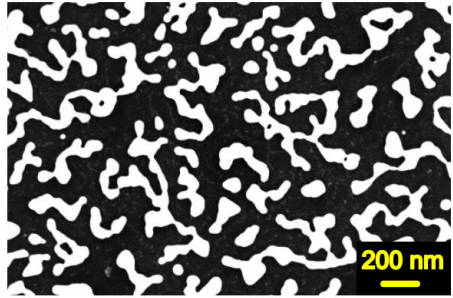

**Supplementary Figure 3.** SEM image of Au nanostructures fabricated by dry synthesis but without using the shadow growth GLAD technique. The Au film on PMMA was deposited under normal incidence without substrate tilting (thickness of 8 nm). The plasma treatment used the same conditions (0.4 mbar, air ambient, 200 W, 15 min) that were used to grow the NPGF. This clearly shows that the resultant film is not continuous in contrast to those seen in Fig. 1c.

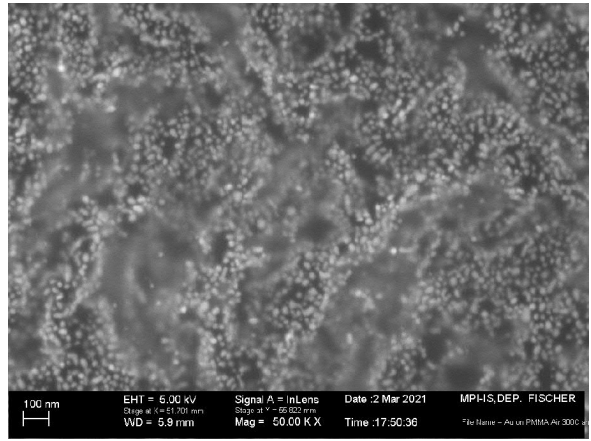

**Supplementary Figure 4**. Au nanoparticles on PMMA with heating at 300 °C for 5 min.

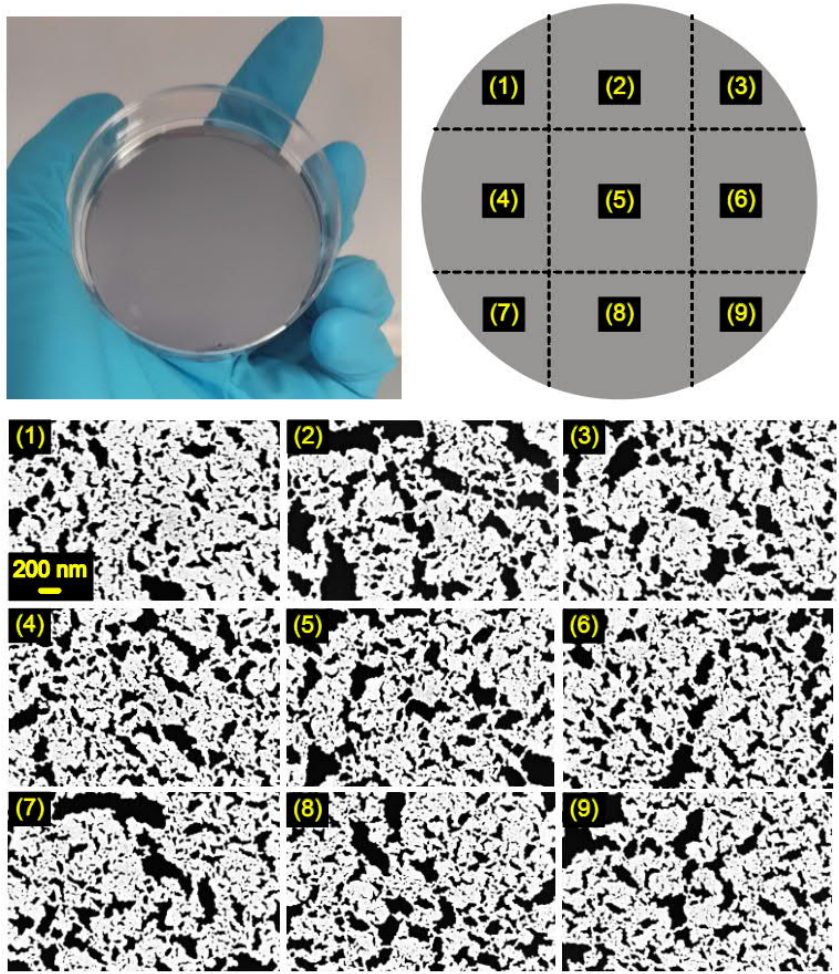

**Supplementary Figure 5.** Nanoporous Pt film fabricated on a 2-inch wafer. The same film morphology is observed across the entire wafer.

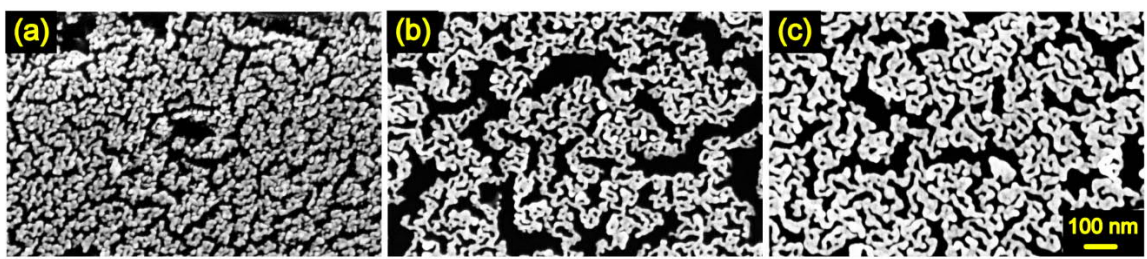

**Supplementary Figure 6**. Morphology changes in the NPGF with varying plasma treatment times. Ligament and pore sizes significantly change because the PMMA is not completely etched. (a) 1min, (b) 5min, and (c) 10 min, 0.4 mbar air ambient with 200 W.

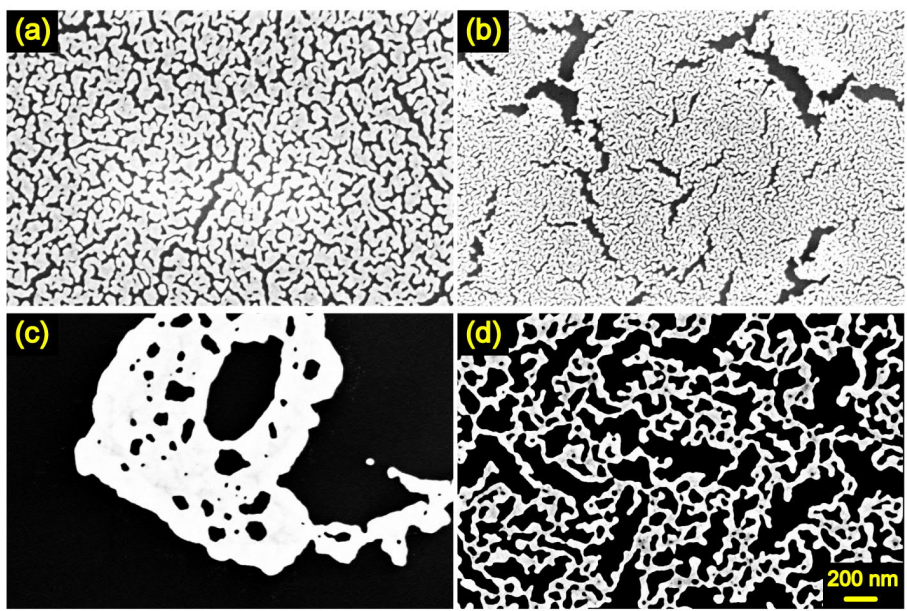

**Supplementary Figure 7**. Morphology control of NPMFs as a function of the plasma conditions. (a-b) NPNFs for different plasma conditions: (a) 0.4 mbar W10 (Ar 90 %, $H_2$ 10 %) ambient with 300 W for 15 min, and (b) 0.4 mbar Ar ambient with 200 W for 15 min. (c-d) NPGF for different plasma conditions: (c) 0.4 mbar W10 (Ar 90 %, $H_2$ 10 %) ambient with 300 W for 15 min, and (d) 0.4 mbar air ambient with 200 W for 25 min.

**Supplementary Table 1**. Melting points and surface energies of metals.[1]

|  | Ag | Au | Ni | Fe | Pd | Pt |
|---|---|---|---|---|---|---|
| **Melting point (°C)** | 961 | 1063 | 1453 | 1538 | 1555 | 1770 |
| **Surface energy (J/m$^2$)** | 1.2 | 1.3-1.7 | 2.0-2.4 | 2.4 | 1.9-2.2 | 2.3-2.8 |

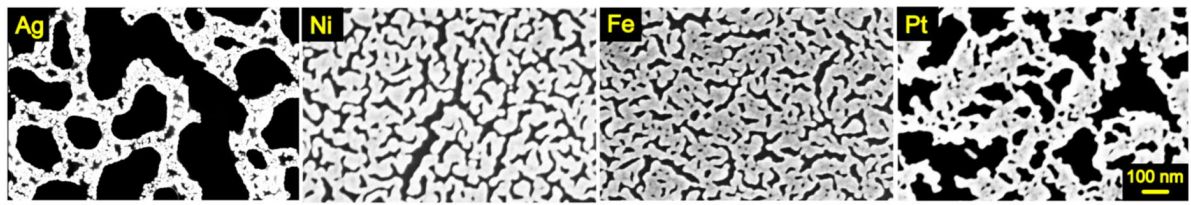

**Supplementary Figure 8.** SEM images of different NPMFs that are fabricated with the same process parameters. Metals were evaporated on a PMMA thin film with an oblique angle of 80 °, with a rate of 0.05 nm/s, 10 nm target thickness, and rotation speed 0.72 °/s. They were plasma-treated in 0.4 mbar W10 ambient (Ar 90 %, $H_2$ 10%) with 300 W for 15 min.

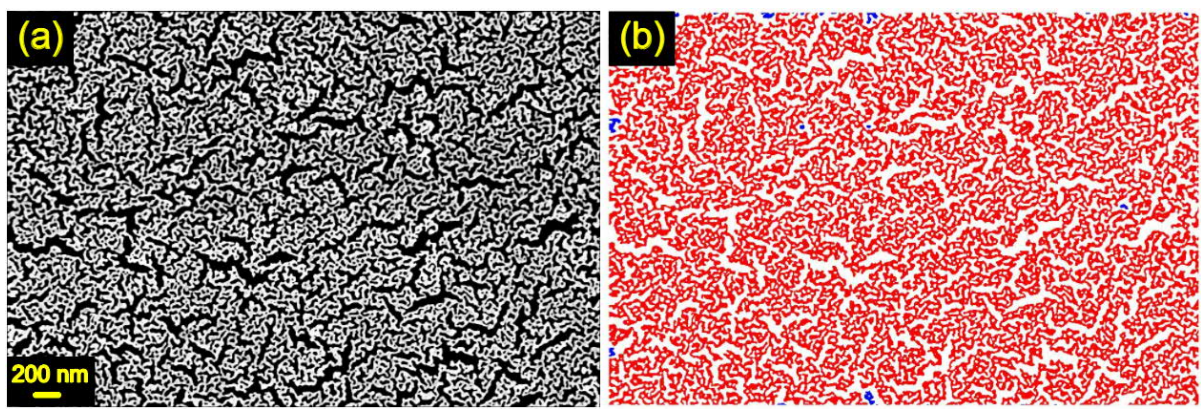

**Supplementary Figure 9.** (a) SEM image of a NPGF. (b) After analyzing the image pixel connectivity (8-connected pixels) of pixels that are connected along the horizontal, vertical, or diagonal direction, connected structures are colored red (unconnected regions are blue and these are almost not present). It is thus seen that almost all ligaments are fully connected. A SEM image of the NPGF film was converted to a binary image with a resolution of 1024 × 680, (black and white) by adapting a thresholding method (the threshold value is 50).[2]

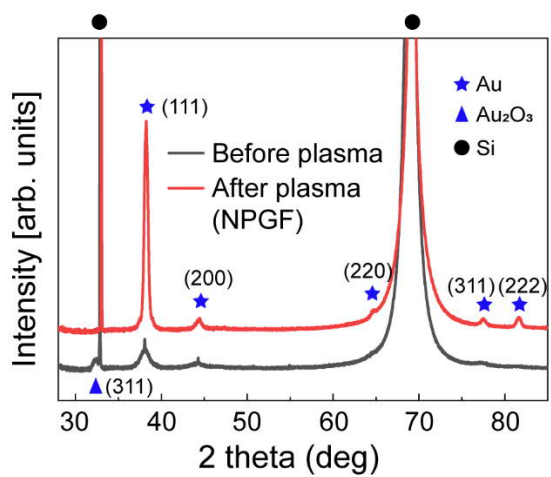

**Supplementary Figure 10**. X-ray diffraction (source: Cu Kα radiation) of NPGF (red) and as-deposited Au on PMMA layer (black) samples. Scan rate is 7 °/h. Peaks from Au are much stronger after plasma treatment. Si peaks are from forbidden (200) reflection and (400).[3] A peak near 32° from the as-deposited Au sample arises $Au_2O_3$, which is naturally formed during the deposition. This peak disappears in the NPGF sample.

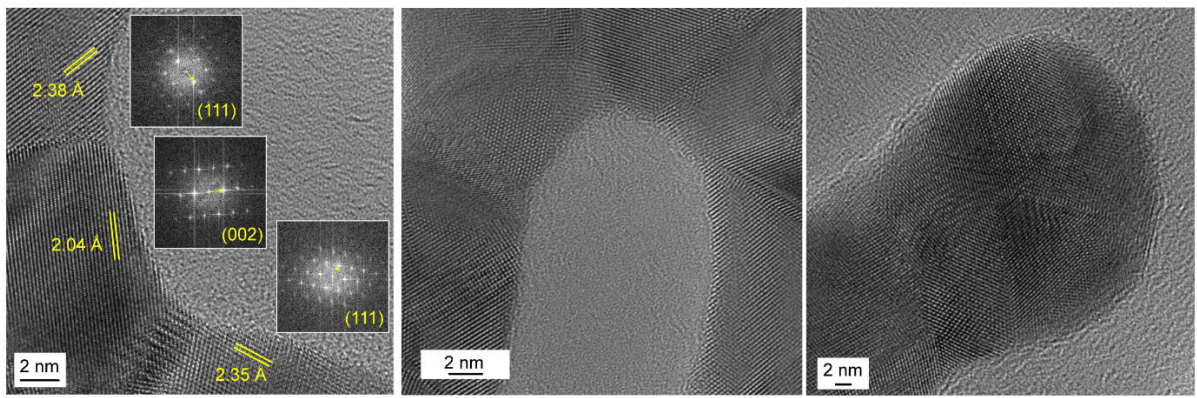

**Supplementary Figure 11**. HR TEM images of concave and protruded regions of a NPGF. Different grains with grain boundaries and a number of different surface facets are clearly seen.

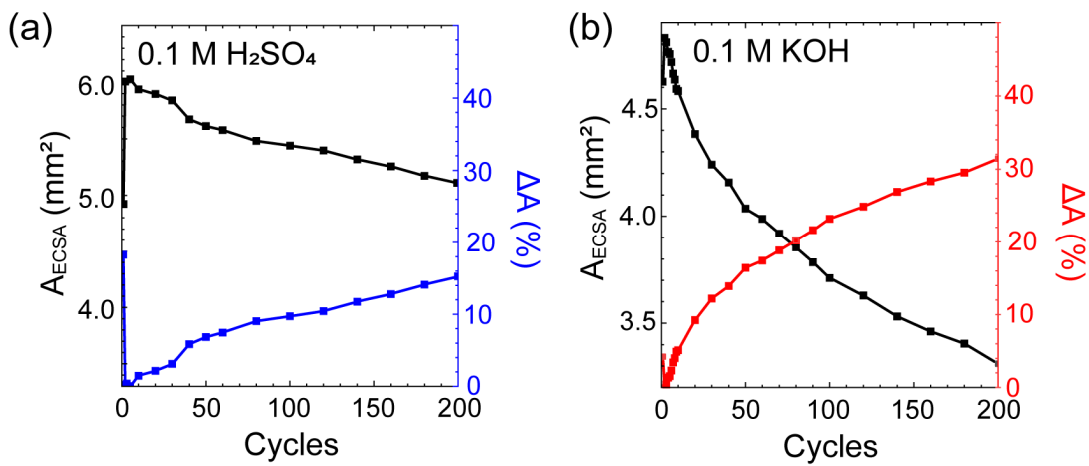

**Supplementary Figure 12.** Estimated electrochemical surface area (ECSA) of NPGF powder during 200 cycles in (a) 0.1 M $H_2SO_4$ and (b) 0.1 M KOH at a scan rate of 10 mV s$^{-1}$. ECSA was quantified by integration of the cathodic peak obtained for the reduction of the previously formed monolayer of gold oxide as depicted in Supplementary Fig. 16. The final ECSA losses after 200 cycles in acidic and alkaline solution were only 16 % and 32 %, respectively.

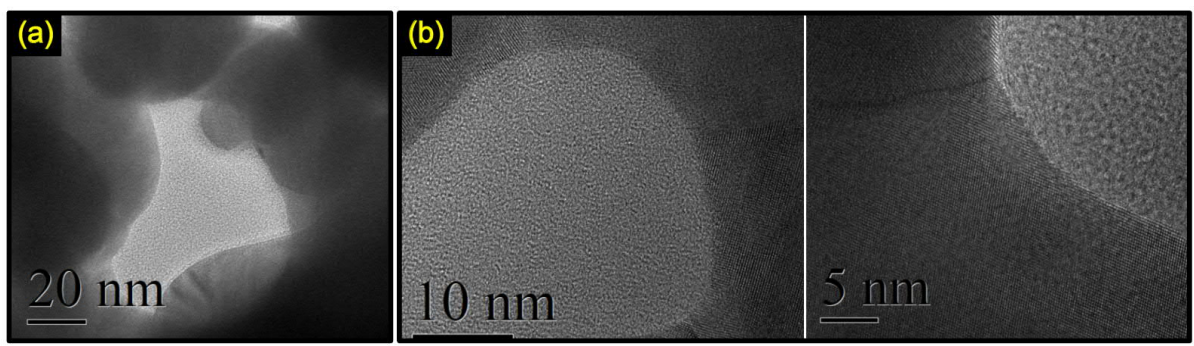

**Supplementary Figure 13.** TEM images of NPGF powder after 200 CV cycles in alkaline media. (a) NPGF powder. (b) HR TEM images of curved structure where grain boundaries are clearly observed. Samples are not as good as one *before* EC measurement, because the removal of the powder from the cavity microelectrode requires extensive sonication which causes fracturing of the NPGF. A single thin layer of Au ligaments was difficult to observe.

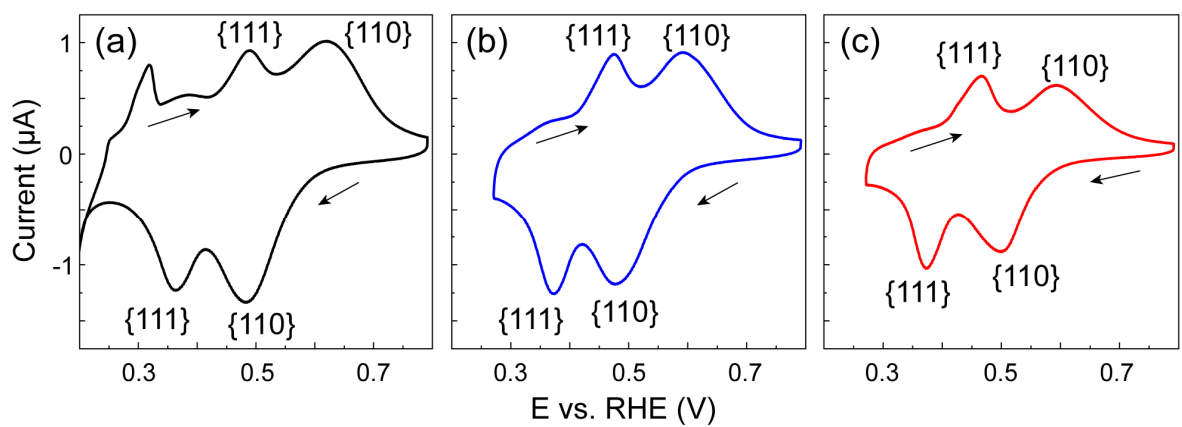

**Supplementary Figure 14.** UPD of Pb on NPGF powder (a) as-prepared, (b) on NPGF powder pre-cycled in acidic solution, and (c) on NPGF powder pre-cycled in alkaline solution. Major peaks during Pb under-potential deposition and dissolution correspond to processes at {111} and {110} domains. Broad peaks are indicative for many defects in NPGF powder. All CVs were measured in 0.1 M KOH + $10^{-3}$ M Pb(NO$_3$)$_2$ at 10 mV s$^{-1}$.

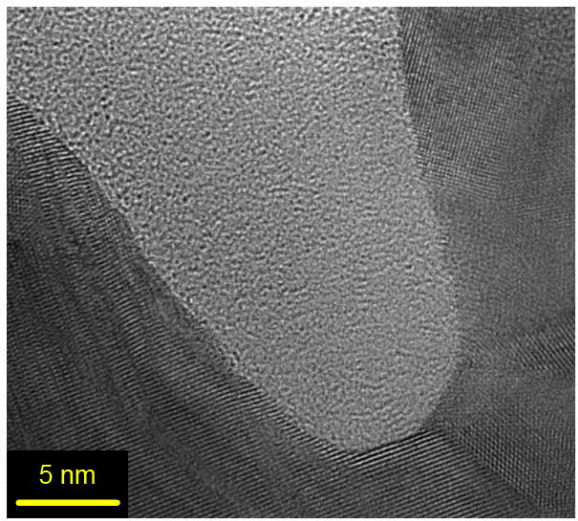

**Supplementary Figure 15.** HR TEM image of the NPGF powder *after* methanol oxidation showing various grains and grain boundaries, and multiple surface directions, indicating the robustness of the NPGF under catalytic conditions.

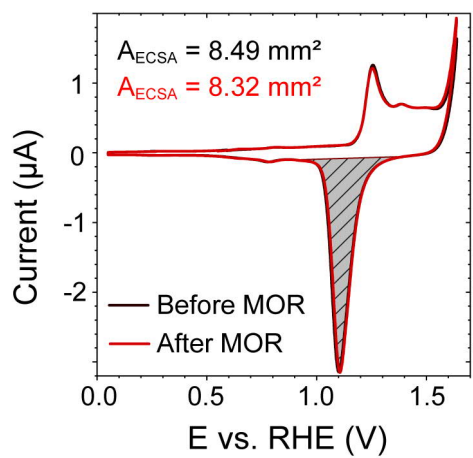

**Supplementary Figure 16.** Cyclic voltammograms of NPGF powder in 0.1 M KOH before (black) and after (red) constant potential MOR at 1.242 V vs. RHE for 2 h. ECSAs before and after MOR were estimated from an integration of the peak (before: gray, after: hatched), showing only a 2% loss.

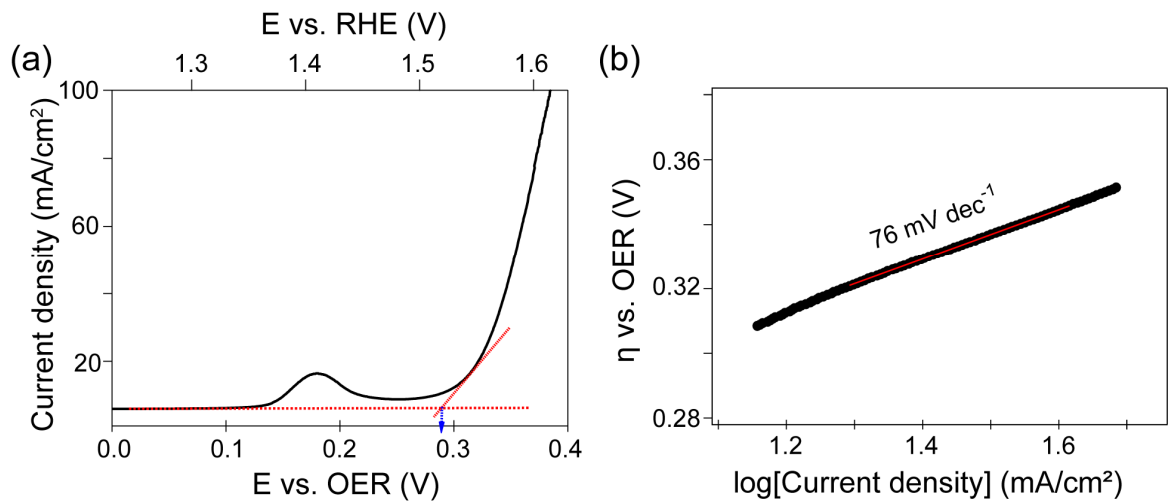

**Supplementary Figure 17.** Electrocatalytic analysis of NPNF powder for OER (a) LSV for the NPNF in 0.1 M KOH at 1 mV s$^{-1}$, showing the fast rise of the current density of the OER achieved at very low overpotentials, onset potential marked with dashed lines. (b) Tafel plot for the OER of NPNF powder, indicating the high activity of the film.

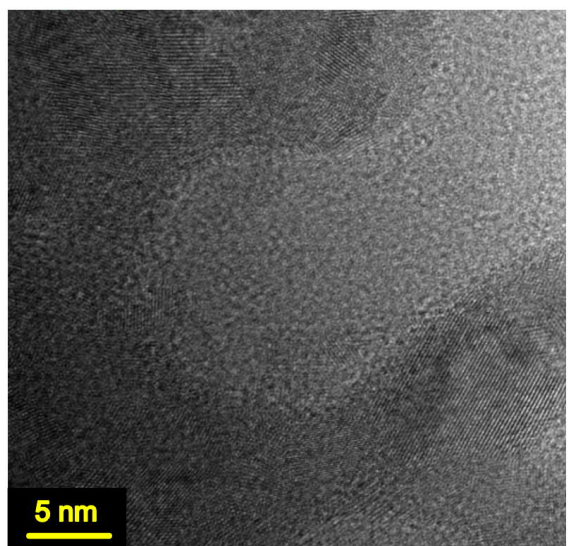

**Supplementary Figure 18.** HR TEM images of NPNF powder. Magnified image of curved region, showing that the NPNF has surfaces with multiple directions in curved regions, and many grains each with sizes of a few nanometers.

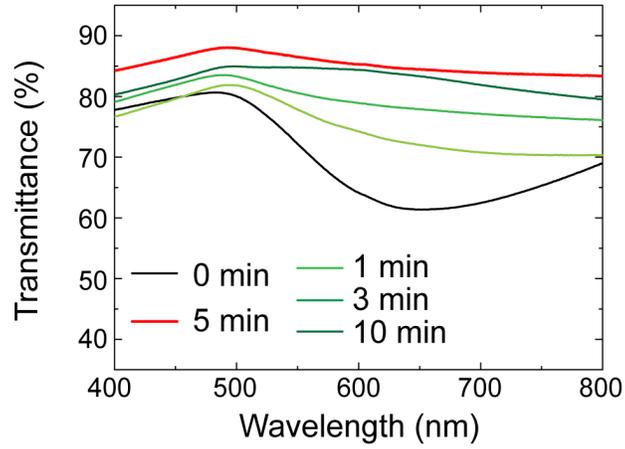

**Supplementary Figure 19**. Measured transmittance of NPGF depending on plasma treatment time.

**Supplementary Table 2**. Measured sheet resistance of NPGF.

| Sample preparation | Sheet resistance |
|---|---|
| No plasma | Not measurable |
| Air, 0.4 mbar, 200 W, 1 min | Not measurable |
| Air, 0.4 mbar, 200 W, 3 min | 270 Ω/sq |
| Air, 0.4 mbar, 200 W, 5 min | 170 Ω/sq |
| Air, 0.4 mbar, 200 W, 10 min | 225 Ω/sq |

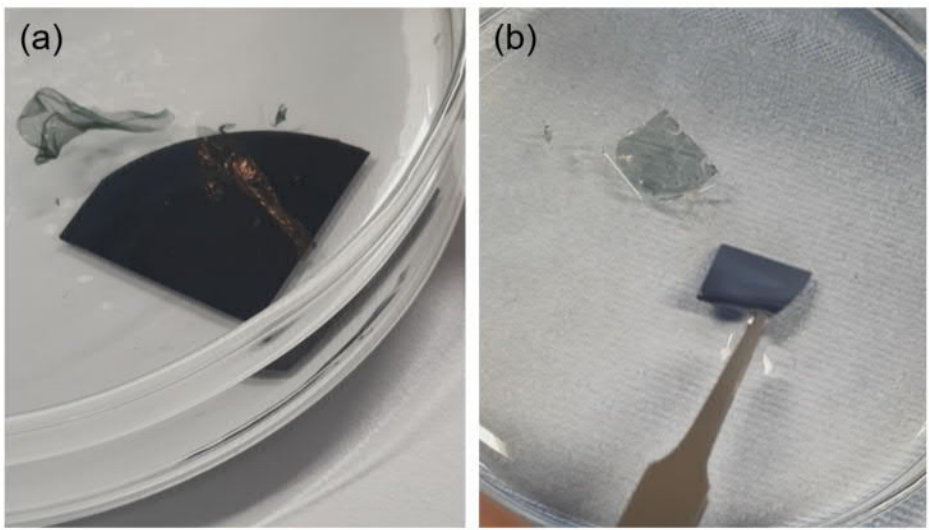

**Supplementary Figure 20**. Free-standing NPGF. (a) Free-standing NPGF in acetone. PMMA that has not been completely etched during the plasma treatment is dissolved in acetone, and NPGF was lifted off from the Si substrate. (b) 3 μm-thick PMMA was spin-coated on NPGF on Si substrate. This sample was immersed in KOH solution (28.5 %) for a minute and rinsed in deionized water, repeatedly, until the film completely lifts-off. Spin-coated PMMA serves as a supporting layer to stabilize NPGF.

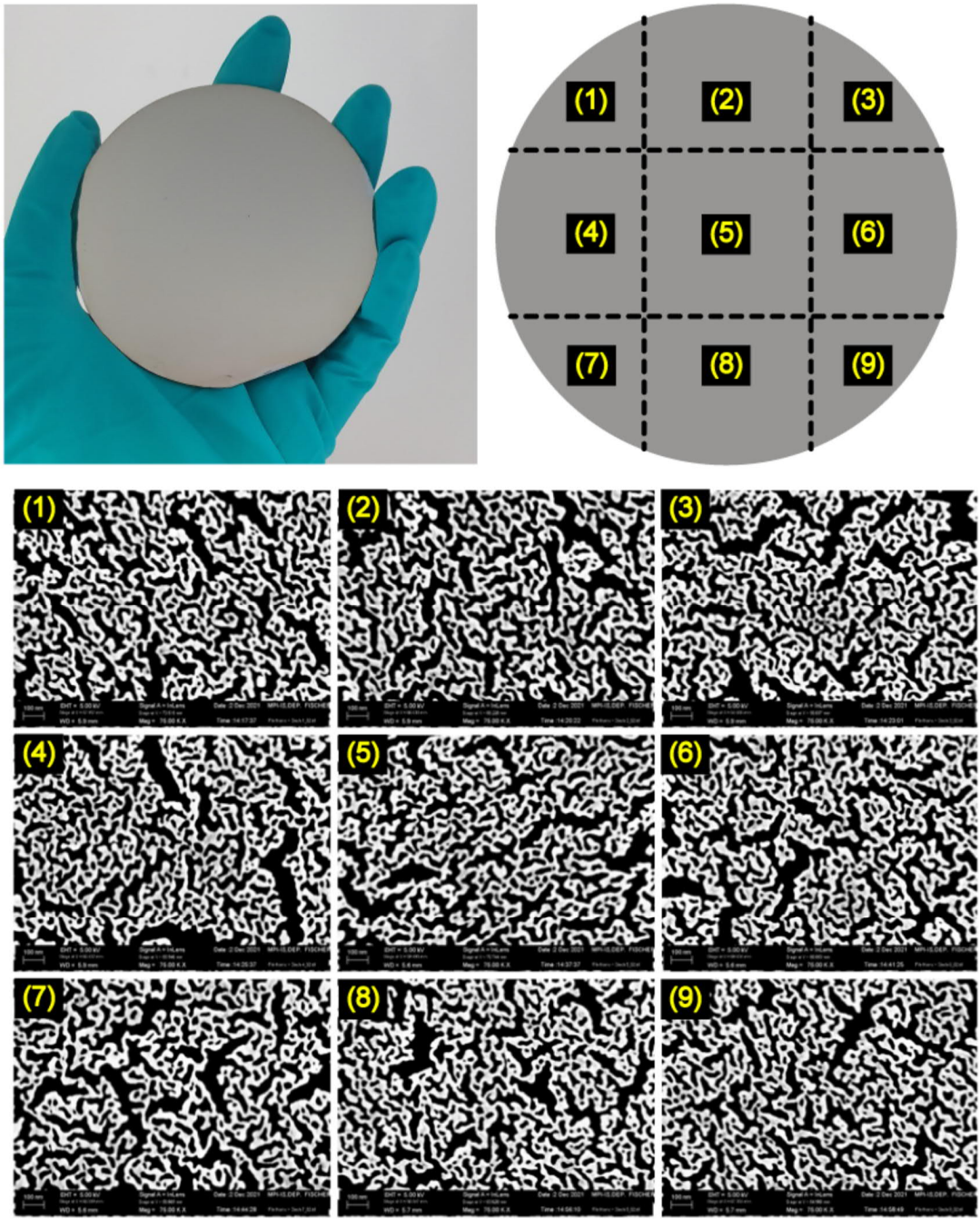

**Supplementary Figure 21.** The NPGF fabricated on 3-inch wafer. SEM images of 9 different regions on the wafer confirm that the NPGF forms on large areas and is uniform across the entire wafer.

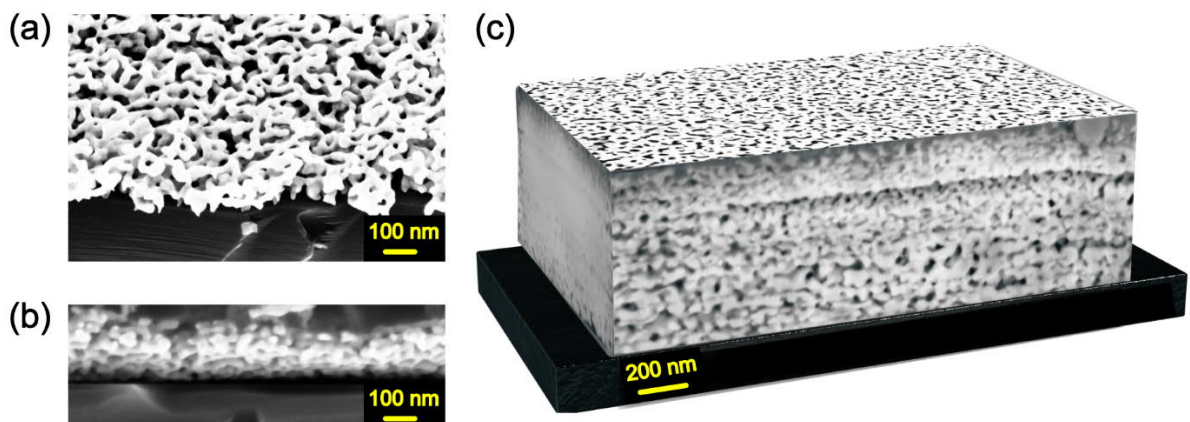

**Supplementary Figure 22.** Multi-stacks of NPGFs: (a) SEM image of a view onto a tilted structure, and (b) cross-section of a five-layer thick stack of NPGF. The stack was obtained by repeating the growth steps: PMMA coating onto a previously obtained film → shadow (GLAD) deposition → plasma treatment. The ligament size is around 20~30 nm which is comparable to that within the single layer NPGF. The layers are connected which is attributed to the repeated plasma treatment steps. (c) Larger view of the structure with the view onto the structure as well as the cross-sectional view of the folded multi-stack NPGF.

The five-layer stack was peeled off in KOH (28 %) solution and transferred to deionized water. The stacked layers could then be mechanically folded with tweezers several times and finally transferred to a Si wafer and dried. The cross-section (lower image in c) was obtained after FIB milling. A slight 'blurring effect' in the electron micrograph is observed in the cross-sectional view, which is caused by the FIB milling.[4,5] Nevertheless, the image shows that it is possible to obtain structures with a reasonable thickness to attain bulk-like structures with a thickness that approaches ~ 1 μm.

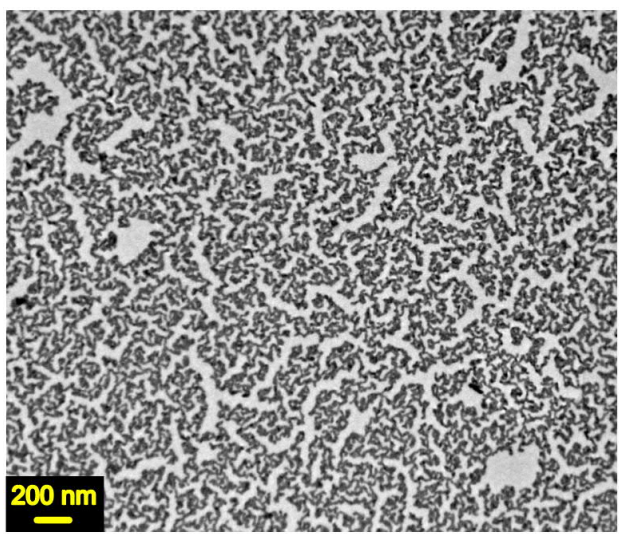

**Supplementary Figure 23.** Low-magnification TEM image of an interconnected NPGF monolayer. The NPGF monolayer was peeled-off in acetone solution, and the floating film was then transferred to a 3.05 mm sized TEM grid.

**References**

1. Vitos, L., Ruban, A. V., Skriver, H. L. & Kollár, J. The surface energy of metals. *Surf. Sci.* **411**, 186–202 (1998).

2. Bradley, D. & Roth, G. Adaptive Thresholding using the Integral Image. *J. Graph. Tools* **12**, 13–21 (2007).

3. Zaumseil, P. High-resolution characterization of the forbidden Si 200 and Si 222 reflections. *J. Appl. Cryst.* **48**, 528–532 (2015).

4. Yang, D., Phillips, N. W., Song, J., Harder, R. J., Cha, W., & Hofmann, F. Annealing of focused ion beam damage in gold microcrystals: an in situ Bragg coherent X-ray diffraction imaging study. Synchrotron Rad. 28, 550-565 (2021).

5. Ananth, M., Stern, L., Ferranti, D., Huynh, C., Notte, J., Scipioni, L., Sanford, C., & Thompson, B. Creating nanohole arrays with the helium ion microscope.